\documentclass[aps,prl,reprint,superscriptaddress]{revtex4-1}

\usepackage{graphicx}
\usepackage{color}

\begin{document}


\title{Comment on ``Pt magnetic polarization on Y$_3$Fe$_5$O$_{12}$ and magnetotransport characteristics''}


\author{Stephan Gepr\"{a}gs}
    \email[]{Stephan.Gepraegs@wmi.badw.de}
    \affiliation{Walther-Mei{\ss}ner-Institut, Bayerische Akademie der Wissenschaften, 85748 Garching, Germany}
    \homepage[]{http://www.wmi.badw.de/}
\author{Sebastian T.B. Goennenwein}
    \affiliation{Walther-Mei{\ss}ner-Institut, Bayerische Akademie der Wissenschaften, 85748 Garching, Germany}
\author{Marc Schneider}
    \affiliation{Walther-Mei{\ss}ner-Institut, Bayerische Akademie der Wissenschaften, 85748 Garching, Germany}
\author{Fabrice Wilhelm}
    \affiliation{European Synchrotron Radiation Facility (ESRF), 38043 Grenoble Cedex 9, France}
\author{Katharina Ollefs}
    \affiliation{European Synchrotron Radiation Facility (ESRF), 38043 Grenoble Cedex 9, France}
\author{Andrei Rogalev}
    \affiliation{European Synchrotron Radiation Facility (ESRF), 38043 Grenoble Cedex 9, France}
\author{Matthias Opel}
    \email[]{Matthias.Opel@wmi.badw.de}
    \affiliation{Walther-Mei{\ss}ner-Institut, Bayerische Akademie der Wissenschaften, 85748 Garching, Germany}
\author{Rudolf Gross}
    \affiliation{Walther-Mei{\ss}ner-Institut, Bayerische Akademie der Wissenschaften, 85748 Garching, Germany}
    \affiliation{Physik-Department, Technische Universit\"{a}t M\"{u}nchen, 85748 Garching, Germany}


\date{\today}



\maketitle



In a recent Letter \cite{Lu2013}, Lu \textit{et al.}~reported on ``ferromagneticlike transport properties'' of thin films of Pt, deposited \textit{ex situ} via sputtering on the ferrimagnetic insulator Y$_3$Fe$_5$O$_{12}$. The authors found a magnetoresistance in Pt displaying a hysteresis corresponding to the coercive field of Y$_3$Fe$_5$O$_{12}$, consistent with the findings of other groups \cite{Nakayama2013,Althammer2013}. While the latter interpreted their data in terms of the recently proposed spin-Hall magnetoresistance \cite{Chen2013}, Lu \textit{et al.}~attributed their observation to a magnetic proximity effect \cite{Lu2013}. To support this interpretation, they measured the X-ray magnetic circular dichroism (XMCD) at the Pt $L_{2,3}$ edges from a Pt/Y$_3$Fe$_5$O$_{12}$ sample with a Pt thickness of $t_\mathrm{Pt}=1.5$~nm and derived an average induced magnetic moment of 0.054 $\mu_{\rm B}$ per Pt atom \cite{Lu2013}. This is contradictory to the results of our previous comprehensive XMCD study of three different Pt/Y$_3$Fe$_5$O$_{12}$ samples ($t_\mathrm{Pt}=3,7,$ and 10\,nm) from which we identified an upper limit of $(0.003 \pm 0.001)\,\mu_{\rm B}$ per Pt \cite{Geprags2012}.

\begin{figure}
  \includegraphics[width=8.5cm]{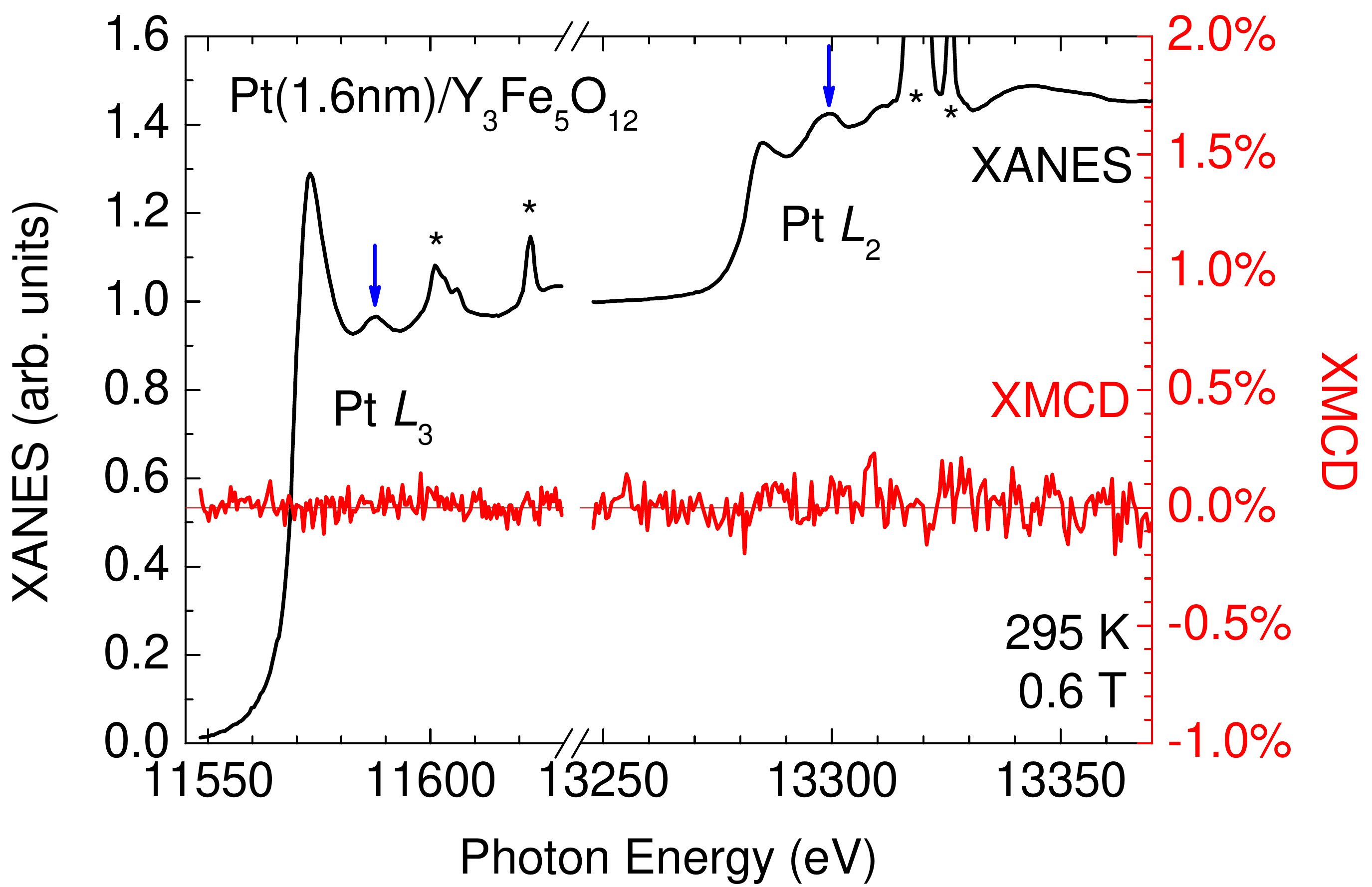}
  \caption{\label{fig}
           Normalized Pt $L_{2,3}$ edge XANES (black, left scale) and XMCD (red, right scale) spectra from Pt/Y$_3$Fe$_5$O$_{12}$ with a Pt thickness of $t_\mathrm{Pt} = 1.6$~nm.
           EXAFS wiggles at 11588\,eV and 13300\,eV are indicated by blue arrows.
           Diffraction peaks are marked by asterisks ($\star$).}
\end{figure}

We here corroborate our statement by a new dataset from a fourth Pt/Y$_3$Fe$_5$O$_{12}$ sample with $t_\mathrm{Pt}=1.6$~nm, \textit{i.e.}~very close to the Pt thickness studied by Lu \textit{et al.}, deposited \textit{in situ} via electron-beam evaporation (Pt) and laser-MBE (Y$_3$Fe$_5$O$_{12}$) on a Y$_3$Al$_5$O$_{12}$ substrate as described earlier \cite{Geprags2012}. The X-ray absorption (XANES) and the corresponding XMCD (Fig.~\ref{fig}) were measured in a magnetic field of $\pm 0.6$\,T parallel to the incoming X-rays under an angle of $\approx 3^\circ$ to the sample surface. The spectra were recorded and normalized as described earlier \cite{Geprags2012}. Our data allow us to exclude a finite XMCD signal at both Pt $L_{2,3}$ edges down to a noise level of $<\,0.1\%$ with respect to the edge jump. This noise level is at least 10 times lower than the XMCD signal of $1\%$ of Lu \textit{et al.}\cite{Lu2013}.

In view of these contradictory XMCD observations, it is instructive to compare the corresponding XANES. The whiteline intensity at the Pt $L_3$ edge is known to represent a sensitive measure for the oxidation state of the absorbing Pt: PtO$_{1.6}$ displays a whiteline intensity of 2.20, PtO$_{1.36}$ of 1.50, and metallic Pt of 1.25 with respect to the edge jump \cite{Kolobov2005}. The intensity of the Pt $L_3$ whiteline in our Pt(1.6nm)/Y$_3$Fe$_5$O$_{12}$ is less than 1.30 (Fig.~\ref{fig}), very close to our previous observations \cite{Geprags2012} and the value reported for metallic Pt \cite{Kolobov2005}. Lu \textit{et al.}, however, find an ``XAS step height of 2.07'' \cite{Lu2013} and a whiteline maximum of roughly 3~a.u.~(Fig.~2(a) in Ref.~\cite{Lu2013}), corresponding to a whiteline intensity of 1.45 relative to the edge jump. This discrepancy is even more pronounced for the Pt $L_2$ whiteline intensities: $0.79$ for Pt metal\cite{Bartolome2009} and $0.80$ for our new dataset, but above unity in Ref.~\cite{Lu2013}. Moreover, we observe EXAFS wiggles at 11588\,eV and 13300\,eV which are characteristic of metallic Pt\cite{Bartolome2009}, but not present in Ref.~\cite{Lu2013}. From this, we deduce that the sample investigated by Lu \textit{et al.}~does not contain a clean Pt metal film, possibly because of partial intermixing with Y$_3$Fe$_5$O$_{12}$ as a consequence of the Pt sputtering process.

In summary, our published \cite{Geprags2012} and new data (Fig.~\ref{fig}) unambiguously show that in our metallic Pt on Y$_3$Fe$_5$O$_{12}$ the induced magnetic moment in Pt is negligible. We do not find any indication for magnetic proximity even in ultrathin layers ($t_\mathrm{Pt} < 2\,$nm) within the accuracy of our XMCD measurement. This calls into question the main conclusion of Lu \textit{et al.}\cite{Lu2013} of a ``magnetic proximity magnetoresistance''. Our data instead strongly support the spin-Hall magnetoresistance model \cite{Nakayama2013,Althammer2013,Chen2013}.

\begin{acknowledgments}
This work was supported by the European Synchrotron Radiation Facility (ESRF) via HE-3784 and the Deutsche Forschungsgemeinschaft (DFG) via SPP 1538 (Project No.~GO 944/4-1).
\end{acknowledgments}

\bibliography{xmcd_comment}

\providecommand{\noopsort}[1]{}\providecommand{\singleletter}[1]{#1}%
\begin{thebibliography}{7}%
\makeatletter
\providecommand \@ifxundefined [1]{%
 \@ifx{#1\undefined}
}%
\providecommand \@ifnum [1]{%
 \ifnum #1\expandafter \@firstoftwo
 \else \expandafter \@secondoftwo
 \fi
}%
\providecommand \@ifx [1]{%
 \ifx #1\expandafter \@firstoftwo
 \else \expandafter \@secondoftwo
 \fi
}%
\providecommand \natexlab [1]{#1}%
\providecommand \enquote  [1]{``#1''}%
\providecommand \bibnamefont  [1]{#1}%
\providecommand \bibfnamefont [1]{#1}%
\providecommand \citenamefont [1]{#1}%
\providecommand \href@noop [0]{\@secondoftwo}%
\providecommand \href [0]{\begingroup \@sanitize@url \@href}%
\providecommand \@href[1]{\@@startlink{#1}\@@href}%
\providecommand \@@href[1]{\endgroup#1\@@endlink}%
\providecommand \@sanitize@url [0]{\catcode `\\12\catcode `\$12\catcode
  `\&12\catcode `\#12\catcode `\^12\catcode `\_12\catcode `\%12\relax}%
\providecommand \@@startlink[1]{}%
\providecommand \@@endlink[0]{}%
\providecommand \url  [0]{\begingroup\@sanitize@url \@url }%
\providecommand \@url [1]{\endgroup\@href {#1}{\urlprefix }}%
\providecommand \urlprefix  [0]{URL }%
\providecommand \Eprint [0]{\href }%
\providecommand \doibase [0]{http://dx.doi.org/}%
\providecommand \selectlanguage [0]{\@gobble}%
\providecommand \bibinfo  [0]{\@secondoftwo}%
\providecommand \bibfield  [0]{\@secondoftwo}%
\providecommand \translation [1]{[#1]}%
\providecommand \BibitemOpen [0]{}%
\providecommand \bibitemStop [0]{}%
\providecommand \bibitemNoStop [0]{.\EOS\space}%
\providecommand \EOS [0]{\spacefactor3000\relax}%
\providecommand \BibitemShut  [1]{\csname bibitem#1\endcsname}%
\let\auto@bib@innerbib\@empty
\bibitem [{\citenamefont {Lu}\ \emph {et~al.}(2013)\citenamefont {Lu},
  \citenamefont {Choi}, \citenamefont {Ortega}, \citenamefont {Cheng},
  \citenamefont {Cai}, \citenamefont {Huang}, \citenamefont {Sun},\ and\
  \citenamefont {Chien}}]{Lu2013}%
  \BibitemOpen
  \bibfield  {author} {\bibinfo {author} {\bibfnamefont {Y.~M.}\ \bibnamefont
  {Lu}}, \bibinfo {author} {\bibfnamefont {Y.}~\bibnamefont {Choi}}, \bibinfo
  {author} {\bibfnamefont {C.~M.}\ \bibnamefont {Ortega}}, \bibinfo {author}
  {\bibfnamefont {X.~M.}\ \bibnamefont {Cheng}}, \bibinfo {author}
  {\bibfnamefont {J.~W.}\ \bibnamefont {Cai}}, \bibinfo {author} {\bibfnamefont
  {S.~Y.}\ \bibnamefont {Huang}}, \bibinfo {author} {\bibfnamefont
  {L.}~\bibnamefont {Sun}}, \ and\ \bibinfo {author} {\bibfnamefont {C.~L.}\
  \bibnamefont {Chien}},\ }\href {\doibase 10.1103/PhysRevLett.110.147207}
  {\bibfield  {journal} {\bibinfo  {journal} {Phys. Rev. Lett.}\ }\textbf
  {\bibinfo {volume} {110}},\ \bibinfo {pages} {147207} (\bibinfo {year}
  {2013})}\BibitemShut {NoStop}%
\bibitem [{\citenamefont {Nakayama}\ \emph {et~al.}(2013)\citenamefont
  {Nakayama}, \citenamefont {Althammer}, \citenamefont {Chen}, \citenamefont
  {Uchida}, \citenamefont {Kajiwara}, \citenamefont {Kikuchi}, \citenamefont
  {Ohtani}, \citenamefont {Gepr\"ags}, \citenamefont {Opel}, \citenamefont
  {Takahashi}, \citenamefont {Gross}, \citenamefont {Bauer}, \citenamefont
  {Goennenwein},\ and\ \citenamefont {Saitoh}}]{Nakayama2013}%
  \BibitemOpen
  \bibfield  {author} {\bibinfo {author} {\bibfnamefont {H.}~\bibnamefont
  {Nakayama}}, \bibinfo {author} {\bibfnamefont {M.}~\bibnamefont {Althammer}},
  \bibinfo {author} {\bibfnamefont {Y.-T.}\ \bibnamefont {Chen}}, \bibinfo
  {author} {\bibfnamefont {K.}~\bibnamefont {Uchida}}, \bibinfo {author}
  {\bibfnamefont {Y.}~\bibnamefont {Kajiwara}}, \bibinfo {author}
  {\bibfnamefont {D.}~\bibnamefont {Kikuchi}}, \bibinfo {author} {\bibfnamefont
  {T.}~\bibnamefont {Ohtani}}, \bibinfo {author} {\bibfnamefont
  {S.}~\bibnamefont {Gepr\"ags}}, \bibinfo {author} {\bibfnamefont
  {M.}~\bibnamefont {Opel}}, \bibinfo {author} {\bibfnamefont {S.}~\bibnamefont
  {Takahashi}}, \bibinfo {author} {\bibfnamefont {R.}~\bibnamefont {Gross}},
  \bibinfo {author} {\bibfnamefont {G.~E.~W.}\ \bibnamefont {Bauer}}, \bibinfo
  {author} {\bibfnamefont {S.~T.~B.}\ \bibnamefont {Goennenwein}}, \ and\
  \bibinfo {author} {\bibfnamefont {E.}~\bibnamefont {Saitoh}},\ }\href
  {\doibase 10.1103/PhysRevLett.110.206601} {\bibfield  {journal} {\bibinfo
  {journal} {Phys. Rev. Lett.}\ }\textbf {\bibinfo {volume} {110}},\ \bibinfo
  {pages} {206601} (\bibinfo {year} {2013})}\BibitemShut {NoStop}%
\bibitem [{\citenamefont {Althammer}\ \emph {et~al.}(2013)\citenamefont
  {Althammer}, \citenamefont {Meyer}, \citenamefont {Nakayama}, \citenamefont
  {Schreier}, \citenamefont {Altmannshofer}, \citenamefont {Weiler},
  \citenamefont {Huebl}, \citenamefont {Gepr\"ags}, \citenamefont {Opel},
  \citenamefont {Gross}, \citenamefont {Meier}, \citenamefont {Klewe},
  \citenamefont {Kuschel}, \citenamefont {Schmalhorst}, \citenamefont {Reiss},
  \citenamefont {Shen}, \citenamefont {Gupta}, \citenamefont {Chen},
  \citenamefont {Bauer}, \citenamefont {Saitoh},\ and\ \citenamefont
  {Goennenwein}}]{Althammer2013}%
  \BibitemOpen
  \bibfield  {author} {\bibinfo {author} {\bibfnamefont {M.}~\bibnamefont
  {Althammer}}, \bibinfo {author} {\bibfnamefont {S.}~\bibnamefont {Meyer}},
  \bibinfo {author} {\bibfnamefont {H.}~\bibnamefont {Nakayama}}, \bibinfo
  {author} {\bibfnamefont {M.}~\bibnamefont {Schreier}}, \bibinfo {author}
  {\bibfnamefont {S.}~\bibnamefont {Altmannshofer}}, \bibinfo {author}
  {\bibfnamefont {M.}~\bibnamefont {Weiler}}, \bibinfo {author} {\bibfnamefont
  {H.}~\bibnamefont {Huebl}}, \bibinfo {author} {\bibfnamefont
  {S.}~\bibnamefont {Gepr\"ags}}, \bibinfo {author} {\bibfnamefont
  {M.}~\bibnamefont {Opel}}, \bibinfo {author} {\bibfnamefont {R.}~\bibnamefont
  {Gross}}, \bibinfo {author} {\bibfnamefont {D.}~\bibnamefont {Meier}},
  \bibinfo {author} {\bibfnamefont {C.}~\bibnamefont {Klewe}}, \bibinfo
  {author} {\bibfnamefont {T.}~\bibnamefont {Kuschel}}, \bibinfo {author}
  {\bibfnamefont {J.-M.}\ \bibnamefont {Schmalhorst}}, \bibinfo {author}
  {\bibfnamefont {G.}~\bibnamefont {Reiss}}, \bibinfo {author} {\bibfnamefont
  {L.}~\bibnamefont {Shen}}, \bibinfo {author} {\bibfnamefont {A.}~\bibnamefont
  {Gupta}}, \bibinfo {author} {\bibfnamefont {Y.-T.}\ \bibnamefont {Chen}},
  \bibinfo {author} {\bibfnamefont {G.~E.~W.}\ \bibnamefont {Bauer}}, \bibinfo
  {author} {\bibfnamefont {E.}~\bibnamefont {Saitoh}}, \ and\ \bibinfo {author}
  {\bibfnamefont {S.~T.~B.}\ \bibnamefont {Goennenwein}},\ }\href {\doibase
  10.1103/PhysRevB.87.224401} {\bibfield  {journal} {\bibinfo  {journal} {Phys.
  Rev. B}\ }\textbf {\bibinfo {volume} {87}},\ \bibinfo {pages} {224401}
  (\bibinfo {year} {2013})}\BibitemShut {NoStop}%
\bibitem [{\citenamefont {Chen}\ \emph {et~al.}(2013)\citenamefont {Chen},
  \citenamefont {Takahashi}, \citenamefont {Nakayama}, \citenamefont
  {Althammer}, \citenamefont {Goennenwein}, \citenamefont {Saitoh},\ and\
  \citenamefont {Bauer}}]{Chen2013}%
  \BibitemOpen
  \bibfield  {author} {\bibinfo {author} {\bibfnamefont {Y.-T.}\ \bibnamefont
  {Chen}}, \bibinfo {author} {\bibfnamefont {S.}~\bibnamefont {Takahashi}},
  \bibinfo {author} {\bibfnamefont {H.}~\bibnamefont {Nakayama}}, \bibinfo
  {author} {\bibfnamefont {M.}~\bibnamefont {Althammer}}, \bibinfo {author}
  {\bibfnamefont {S.~T.~B.}\ \bibnamefont {Goennenwein}}, \bibinfo {author}
  {\bibfnamefont {E.}~\bibnamefont {Saitoh}}, \ and\ \bibinfo {author}
  {\bibfnamefont {G.~E.~W.}\ \bibnamefont {Bauer}},\ }\href {\doibase
  10.1103/PhysRevB.87.144411} {\bibfield  {journal} {\bibinfo  {journal} {Phys.
  Rev. B}\ }\textbf {\bibinfo {volume} {87}},\ \bibinfo {pages} {144411}
  (\bibinfo {year} {2013})}\BibitemShut {NoStop}%
\bibitem [{\citenamefont {Gepr\"{a}gs}\ \emph {et~al.}(2012)\citenamefont
  {Gepr\"{a}gs}, \citenamefont {Meyer}, \citenamefont {Altmannshofer},
  \citenamefont {Opel}, \citenamefont {Wilhelm}, \citenamefont {Rogalev},
  \citenamefont {Gross},\ and\ \citenamefont {Goennenwein}}]{Geprags2012}%
  \BibitemOpen
  \bibfield  {author} {\bibinfo {author} {\bibfnamefont {S.}~\bibnamefont
  {Gepr\"{a}gs}}, \bibinfo {author} {\bibfnamefont {S.}~\bibnamefont {Meyer}},
  \bibinfo {author} {\bibfnamefont {S.}~\bibnamefont {Altmannshofer}}, \bibinfo
  {author} {\bibfnamefont {M.}~\bibnamefont {Opel}}, \bibinfo {author}
  {\bibfnamefont {F.}~\bibnamefont {Wilhelm}}, \bibinfo {author} {\bibfnamefont
  {A.}~\bibnamefont {Rogalev}}, \bibinfo {author} {\bibfnamefont
  {R.}~\bibnamefont {Gross}}, \ and\ \bibinfo {author} {\bibfnamefont
  {S.~T.~B.}\ \bibnamefont {Goennenwein}},\ }\href {\doibase 10.1063/1.4773509}
  {\bibfield  {journal} {\bibinfo  {journal} {Applied Physics Letters}\
  }\textbf {\bibinfo {volume} {101}},\ \bibinfo {eid} {262407} (\bibinfo {year}
  {2012})}\BibitemShut {NoStop}%
\bibitem [{\citenamefont {Kolobov}\ \emph {et~al.}(2005)\citenamefont
  {Kolobov}, \citenamefont {Wilhelm}, \citenamefont {Rogalev}, \citenamefont
  {Shima},\ and\ \citenamefont {Tominaga}}]{Kolobov2005}%
  \BibitemOpen
  \bibfield  {author} {\bibinfo {author} {\bibfnamefont {A.~V.}\ \bibnamefont
  {Kolobov}}, \bibinfo {author} {\bibfnamefont {F.}~\bibnamefont {Wilhelm}},
  \bibinfo {author} {\bibfnamefont {A.}~\bibnamefont {Rogalev}}, \bibinfo
  {author} {\bibfnamefont {T.}~\bibnamefont {Shima}}, \ and\ \bibinfo {author}
  {\bibfnamefont {J.}~\bibnamefont {Tominaga}},\ }\href {\doibase
  10.1063/1.1886255} {\bibfield  {journal} {\bibinfo  {journal} {Applied
  Physics Letters}\ }\textbf {\bibinfo {volume} {86}},\ \bibinfo {eid} {121909}
  (\bibinfo {year} {2005})}\BibitemShut {NoStop}%
\bibitem [{\citenamefont {Bartolom\'{e}}\ \emph {et~al.}(2009)\citenamefont
  {Bartolom\'{e}}, \citenamefont {Bartolom\'{e}}, \citenamefont {Garc\'{\i}a},
  \citenamefont {Roduner}, \citenamefont {Akdogan}, \citenamefont {Wilhelm},\
  and\ \citenamefont {Rogalev}}]{Bartolome2009}%
  \BibitemOpen
  \bibfield  {author} {\bibinfo {author} {\bibfnamefont {J.}~\bibnamefont
  {Bartolom\'{e}}}, \bibinfo {author} {\bibfnamefont {F.}~\bibnamefont
  {Bartolom\'{e}}}, \bibinfo {author} {\bibfnamefont {L.~M.}\ \bibnamefont
  {Garc\'{\i}a}}, \bibinfo {author} {\bibfnamefont {E.}~\bibnamefont
  {Roduner}}, \bibinfo {author} {\bibfnamefont {Y.}~\bibnamefont {Akdogan}},
  \bibinfo {author} {\bibfnamefont {F.}~\bibnamefont {Wilhelm}}, \ and\
  \bibinfo {author} {\bibfnamefont {A.}~\bibnamefont {Rogalev}},\ }\href
  {\doibase 10.1103/PhysRevB.80.014404} {\bibfield  {journal} {\bibinfo
  {journal} {Phys. Rev. B}\ }\textbf {\bibinfo {volume} {80}},\ \bibinfo
  {pages} {014404} (\bibinfo {year} {2009})}\BibitemShut {NoStop}%
\end{thebibliography}%

\end{document}